\documentclass{iaus}
\usepackage{graphicx}

\newcommand{\HI}{{\sc H\,i}}

\title[WSRT HALOGAS Survey]
{The Westerbork HALOGAS Survey:\\Status and Early Results}

\author[G. Heald et al.]
{George Heald$^1$, John Allan$^2$, Laura Zschaechner$^3$, Peter Kamphuis$^4$, Rich Rand$^3$, Gyula J\'ozsa$^1$, \and Gianfranco Gentile$^5$}

\affiliation{$^1$ASTRON, Postbus 2, 7990 AA Dwingeloo, the Netherlands \\
email: {\tt heald@astron.nl} \\[\affilskip]
$^2$Department of Physics and Astronomy, Macalaster College, 1600 Grand Ave, Saint Paul, MN 55105, USA\\[\affilskip]
$^3$Department of Physics and Astronomy, University of New Mexico, 1919 Lomas Blvd NE, Albuquerque, NM 87131, USA\\[\affilskip]
$^4$Astronomisches Institut der Ruhr-Universit\"at Bochum, Universit\"atsstra{\ss}e 150, 44780 Bochum, Germany\\[\affilskip]
$^5$Sterrenkundig Observatorium, Ghent University, Krijgslaan 281, S9, 9000 Ghent, Belgium}

\pubyear{2011}
\volume{277}
\pagerange{NNN}
\jname{Tracing the Ancestry of Galaxies on the Land of our Ancestors}
\editors{C. Carignan, K. Freeman \& F. Combes eds.}
\begin{document}

\maketitle

\begin{abstract}
We present early results from the ongoing Hydrogen Accretion in LOcal GAlaxieS (HALOGAS) Survey, which is being performed with the Westerbork Synthesis Radio Telescope (WSRT). The HALOGAS Survey aims to detect and characterize the cold gas accretion process in nearby spirals, through sensitive observations of neutral hydrogen (\HI) 21-cm line emission. In this contribution, we present an overview of ongoing analyses of several HALOGAS targets.
\end{abstract}

\firstsection 
\section{Survey description and status}\label{section:intro}

The scientific motivation for the HALOGAS Survey, and a detailed description of the observational setup, are provided by \cite{heald_etal_2011}. Briefly, we have selected a sample of 24 edge-on and moderately inclined nearby galaxies\footnote{Two of the 24 HALOGAS sample galaxies, NGC 891 and NGC 2403, have already been the subject of deep \HI\ observations and are not reobserved in our program.}, using neutral criteria, for deep (120 hr) \HI\ line observations at WSRT. The observations are sensitive to faint, diffuse gas (typical column density sensitivity $\sim10^{19}\,\mathrm{cm^{-2}}$) and to small unresolved \HI\ clouds (typical mass sensitivity of order $10^5\,M_{\odot}$). Within the HALOGAS project, each galaxy will be carefully studied to ascertain whether signs of gas accretion are present. Results from the individual targets will be combined to give the first indication of the ubiquity and general characteristics of the cold gas accretion process in nearby spirals.

At the time of writing, the \HI\ observations are nearly completed; we anticipate that the guaranteed WSRT observations (20 of 22 targets) will be finished before summer 2011. A number of ancillary programs are also in progress and will supplement the neutral gas observations: deep UV (GALEX) and optical (INT) images are being obtained, and a novel multi-slit technique will obtain 3D optical spectroscopy of the ionized gas in the edge-on survey targets (\cite{wu_etal_2011}).

\section{Early results}

First results, based on the first semester of survey observations (the HALOGAS Pilot Survey), are presented by \cite{heald_etal_2011}. The survey targets presented in that work are UGC 2082, NGC 672, NGC 925, and NGC 4565. In this contribution we present preliminary analysis for three additional survey targets. A more complete analysis for each of these is in preparation.

\subsection{NGC 1003}\label{subsection:NGC1003}

\begin{figure}[t]
\begin{center}
 \includegraphics[width=0.25\textwidth]{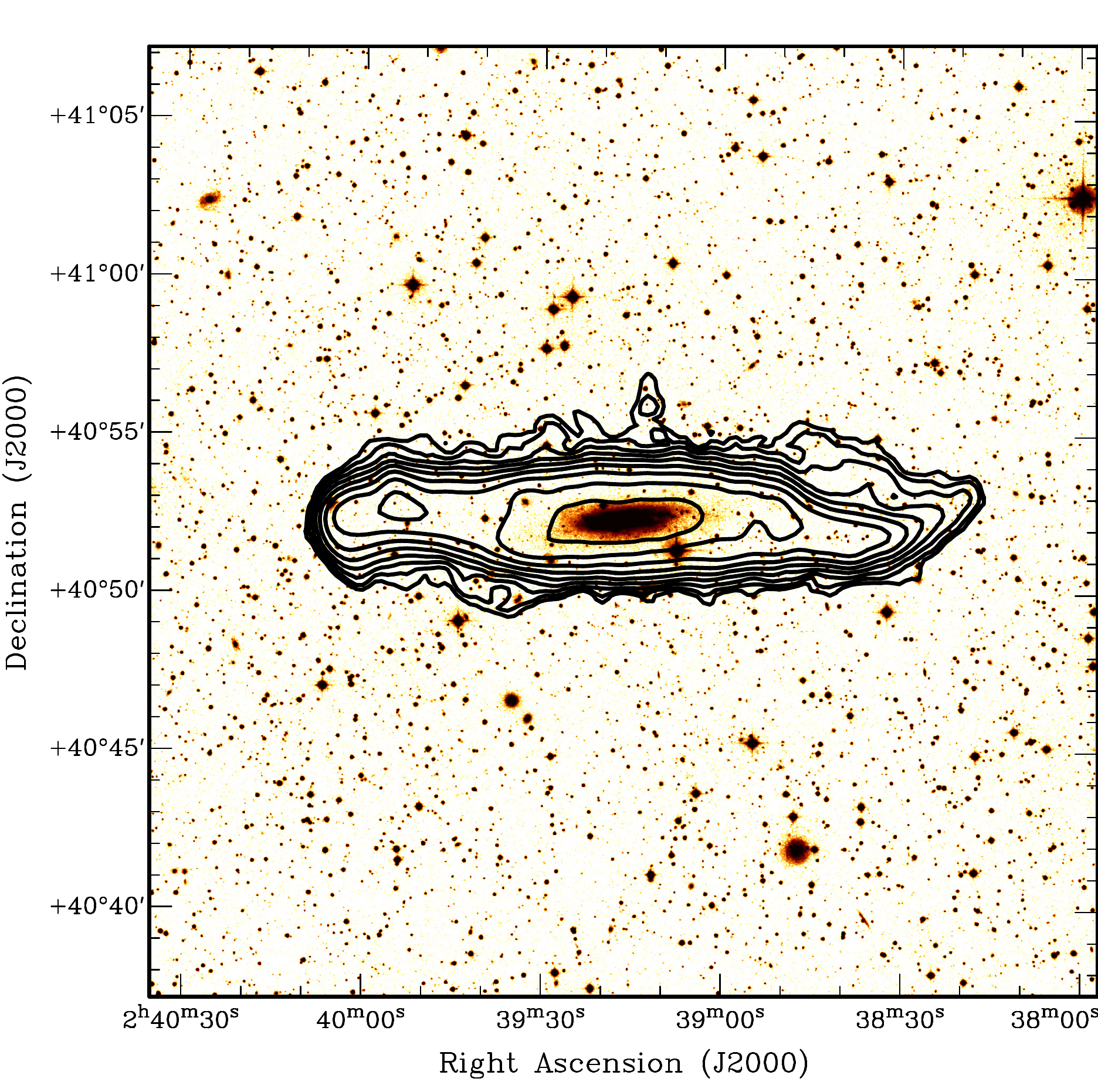}\includegraphics[width=0.25\textwidth]{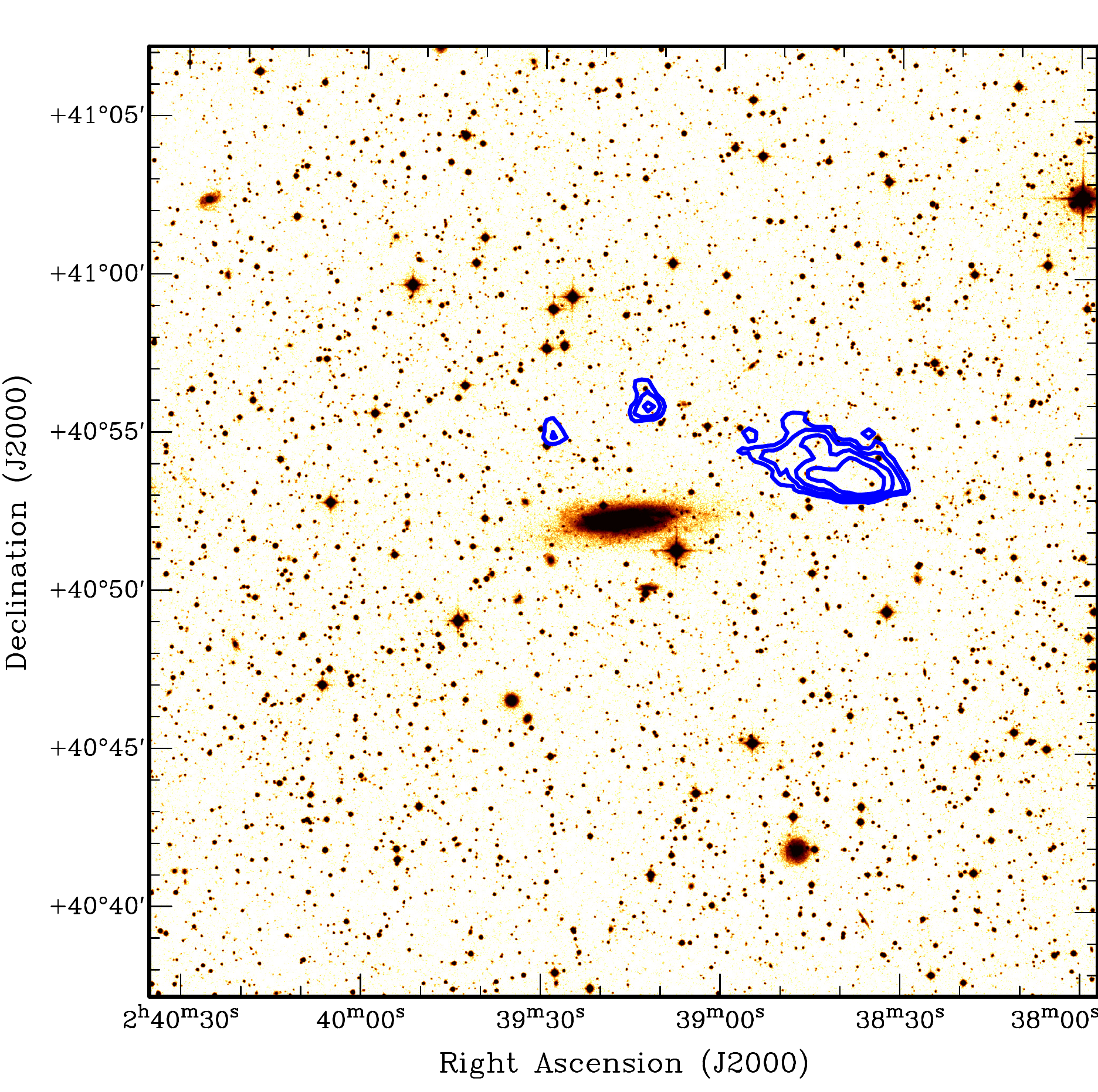}\includegraphics[width=0.5\textwidth]{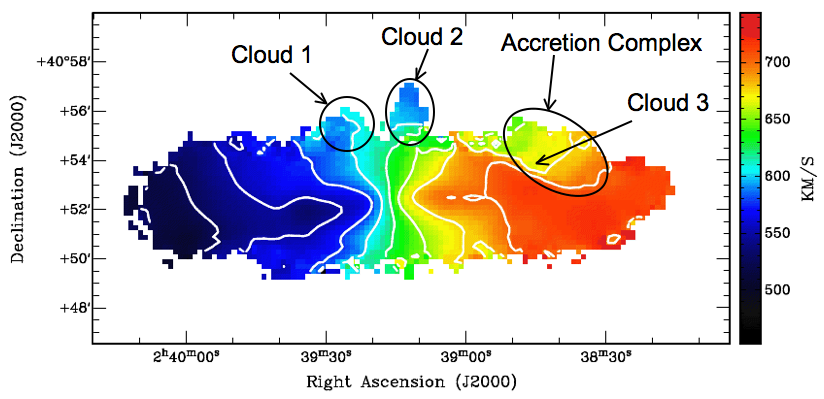} 
 \caption{Neutral hydrogen in NGC 1003. Left: Total \HI\ (moment-0) map overlaid on DSS image. Center: Total \HI\ (moment-0) map constructed from HVC analog \HI\ emission only. Right: Velocity field (moment-1 map), with cloud positions indicated.}
   \label{Figure:NGC1003}
\end{center}
\end{figure}

NGC 1003 has a star formation rate of $0.34\,M_\odot\,\mathrm{yr}^{-1}$ (\cite{heald_etal_2011}). The width of the \HI\ disk is $21^\prime$, compared to the roughly $5^\prime$ optical disk (see Figure \ref{Figure:NGC1003}). Assuming $D\,=\,11\,\mathrm{Mpc}$, we obtain a total \HI\ mass of $5.34\times10^9\,M_\odot$. Our tilted-ring modeling (described below) indicates a strongly warped, flaring thick disk. Several distinct \HI\ features are detected, as shown in Figure \ref{Figure:NGC1003}. None of the clouds, nor the accretion complex, are visible in the optical DSS plates. Cloud 1 has an \HI\ mass of $2\times10^5\,M_\odot$. Cloud 2 has $3.3\times10^5\,M_\odot$ of \HI. Cloud 3 is an unresolved object within the accretion complex with an \HI\ mass of $3\times10^5\,M_\odot$. The accretion complex, excluding Cloud 3, has an \HI\ mass of $2.9\times10^6\,M_\odot$. The Milky Way and M31 are both known to have intermediate- and high-velocity clouds similar to these. The masses we obtain here, as well as the heights above the disk, are consistent with those derived by \cite{wakker_etal_2008} for Milky Way clouds and complexes. Over a dynamical time, they can provide fresh gas compensating for only $\sim2\%$ of the SFR in NGC 1003.

The GIPSY program {\tt GALMOD} was used to hand-craft a tilted-ring model of NGC 1003. Literature values for scale height, rotational velocity, and other parameters were used, and inclination and position angle were estimated. A radial profile was made and used for surface brightness values. This model was refined until further improvements could no longer be made, and the best parameters were fed into the Tilted Ring Fitting Code (TiRiFiC) developed by \cite{jozsa_etal_2007}. The program fits an arbitrary number of parameters at a time for each of the galaxy's rings. We varied a limited number of parameters at once (typically two to three) and then fixed them at their optimum values. Three final models were produced: a thin-disk model, in which scale height was fixed at around $200\,\mathrm{pc}$; a thick-disk model, in which scale height was varied as a unit for all rings (and resulted in a scale height of $\sim1\,\mathrm{kpc}$), and a flare model where the scale height was varied individually for all rings. The east and west halves of the galaxy were modeled separately because of an asymmetrical ``blob'' feature in the eastern side of the disk.

\subsection{NGC 4244}

\begin{figure}
   \begin{center}
   \includegraphics[width=0.5\textwidth]{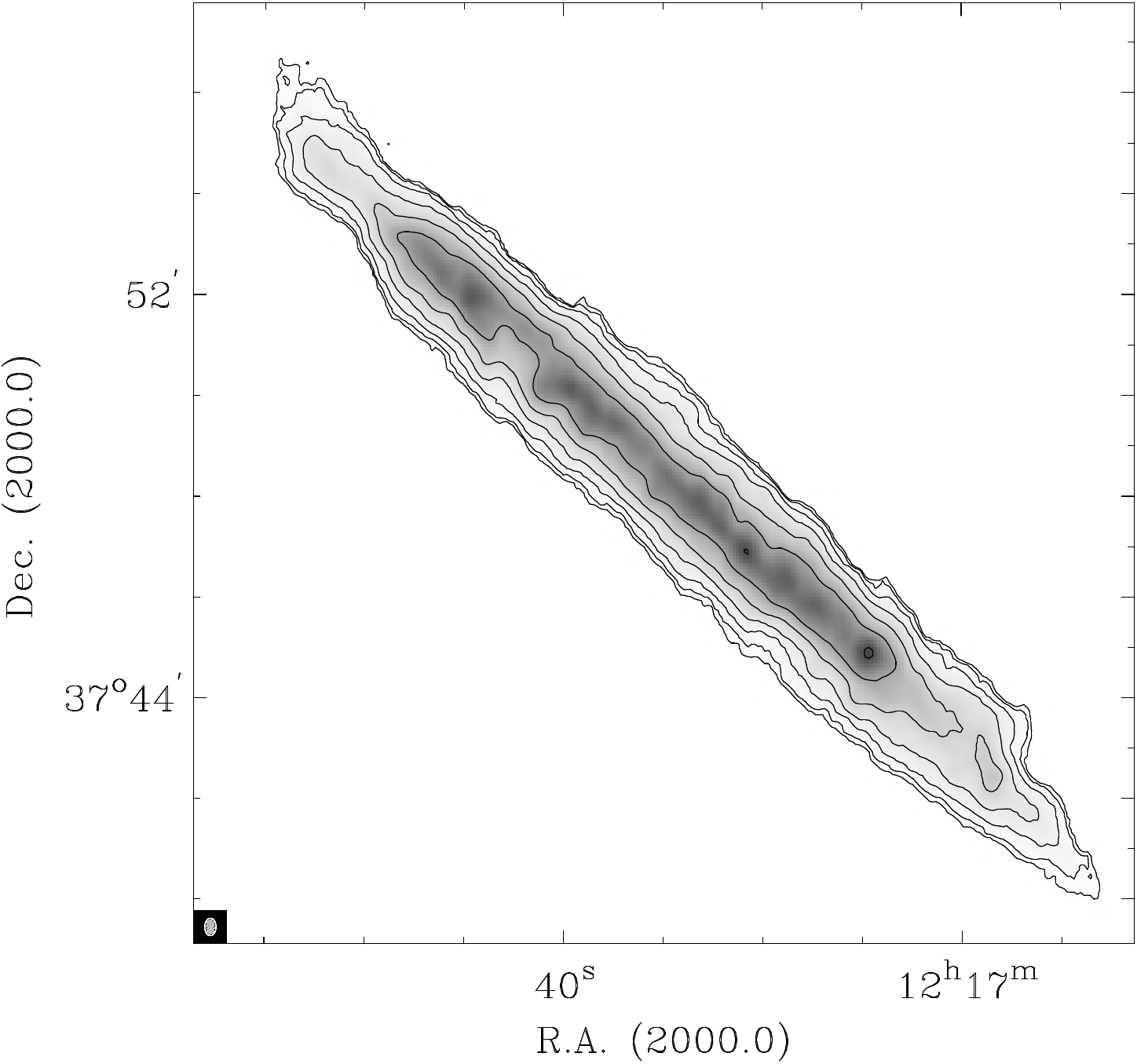}\includegraphics[width=0.5\textwidth]{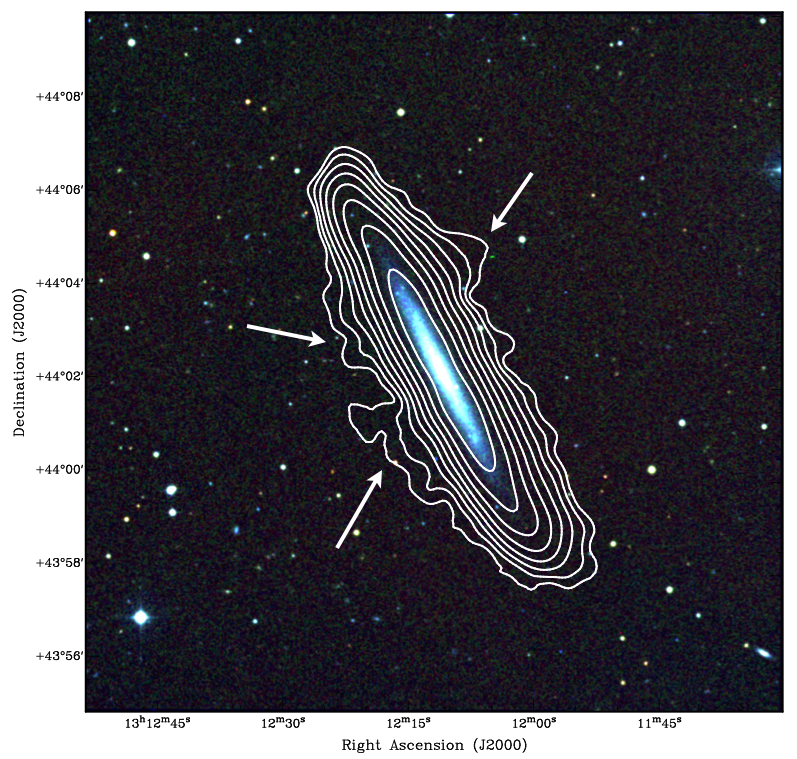} 
   \caption{Left: Integrated \HI\ map of NGC 4244, corrected for primary beam attenuation. Contours begin at $N_{\mathrm{HI}}\,=\,1.4\times10^{20}\,\mathrm{cm}^{-2}$ and increase by factors of two. Right: Integrated \HI\ map of NGC 5023 overlaid on a false-color optical picture constructed from DSS images. The contours begin at $N_{\mathrm{HI}}\,=\,1.9\times10^{19}\,\mathrm{cm}^{-2}$ and increase by factors of two. The arrows indicate individual extraplanar features of interest.}
   \label{Figure:NGC4244_NGC5023}
   \end{center}
\end{figure}

NGC 4244, having a star formation rate of $0.058\,M_{\odot}\,\mathrm{yr}^{-1}$ (\cite{heald_etal_2011}), is on the low star-forming end of the HALOGAS sample. Its high inclination ($88\,^{\circ}$) allows for readily determining the vertical \HI\ structure, as well as an accurate assessment of the kinematics of the neutral gas. For these reasons it will provide substantial insight concerning any connection between star formation and \HI\ halo properties. The total \HI\ map is shown in Figure \ref{Figure:NGC4244_NGC5023}. The disk is clearly warped, and does not show evidence for a bright diffuse halo component. Through careful tilted-ring modeling (similar to that described in \S\,\ref{subsection:NGC1003} but without the use of TiRiFiC), the disk of NGC 4244 is found to be warped both parallel to, as well as perpendicular to, the line of sight. The modeling confirms that we do not detect a substantially extended \HI\ halo. We do, however, detect vertical gradients (lags) in the rotational speed, with magnitudes $-12\,\pm\,2\,\mathrm{km\,s^{-1}\,kpc^{-1}}$ and $-9\,\pm\,2\,\mathrm{km\,s^{-1}\,kpc^{-1}}$ in the approaching and receding halves respectively. These lags decrease in magnitude to $-7\,\pm\,2\,\mathrm{km\,s^{-1}\,kpc^{-1}}$ near a radius of $9\,\mathrm{kpc}$. The detection of the radial variation in the vertical velocity structure will be valuable information for comparison with physical models of the disk-halo interface in spiral galaxies.

Additionally, a few prominent localized features are detected in NGC 4244. Among them is a shell in the approaching half, located directly above a region of star formation. A second feature in the receding half displays an elongated, curved path in the \HI\ contours, extending away from the major axis to approximately $1.5^\prime$ above the midplane. This feature also appears to correspond with star formation, exhibiting a possible hole or vent in the \HI, which is partially filled by H$\alpha$ emission (\cite{hoopes_etal_1999}), indicating a possible connection between the two. Finally, a pronged, faint streak is seen over a velocity range of $\approx\,50\,\mathrm{km\,s^{-1}}$ (cf. the rotation speed, $95-100\,\mathrm{km\,s}^{-1}$) in the receding half. This feature shows no apparent connection to star formation and exhibits traits similar to ram pressure stripping.

\subsection{NGC 5023}

The edge-on galaxy NGC 5023 is a small ($v_{\mathrm{rot}}\,\sim\,80\,\mathrm{km\,s}^{-1}$), slightly warped galaxy at a distance of $6.6\,\mathrm{Mpc}$ (\cite{heald_etal_2011}).  Figure \ref{Figure:NGC4244_NGC5023} shows an integrated \HI\ (moment-0) map. The 21-cm line emission in this galaxy extends on average up to a projected distance of $\sim\,80^{\prime\prime}$ from the plane, with individual features extending to $\sim\,100^{\prime\prime}$ (indicated in the figure with arrows). At the adopted distance, these vertical distances translate to 2.6 kpc and 3.2 kpc, respectively. The most vertically extended features also clearly show up in the individual channel maps of the data. Even though the bulk of the high-latitude emission could be projected above the major axis due to a line-of-sight warp, it is more difficult to explain the individual features in this manner. The extraplanar gas also has lower projected rotational velocities than the gas in the mid-plane. Detailed modeling is in progress to show whether the kinematics is due to a line-of sight warp, a vertical gradient in the rotation curve (as seen in e.g. NGC 4244), or some other effect.

The star formation rate in NGC 5023 is extremely low ($0.032\,M_\odot\,\mathrm{yr}^{-1}$; \cite{heald_etal_2011}). Attributing the extended extraplanar \HI\ features to galactic fountain activity would therefore seem unlikely. However, deep H$\alpha$ imaging performed by \cite{rand_1996} shows faint diffuse {\it ionized} emission in the extraplanar regions -- at the same radial locations as the \HI\ features indicated in Figure \ref{Figure:NGC4244_NGC5023} (they are however not detected to the same vertical distances as the \HI\ features). The correlation between extraplanar diffuse ionized gas (EDIG) features and star formation has been well established (e.g. \cite{rand_1996}). This may indicate that the extraplanar \HI\ features in NGC 5023 are indeed the result of star formation activity. Our 3D optical spectroscopic observations (mentioned in \S\,\ref{section:intro}) of this target will be very interesting for comparison with the \HI\ data.

\section{Intermediate conclusions and ongoing work}

As the HALOGAS Survey progresses, the number of galaxies with well-studied \HI\ morphology and kinematics continues to grow. The extraplanar \HI\ features detected in HALOGAS targets have so far been more subtle than, e.g., the massive multiphase gaseous halo in NGC 891. However, it is not yet clear which characteristics of the underlying galaxy dictate the gaseous content of spiral galaxy halos. It does not seem to be the case that halos are tied in a straightforward way to the SFR of the host galaxy. NGC 4244, for example, has a low SFR and no halo. As shown here, NGC 5023, with a similarly low SFR, does have extended extraplanar \HI\ features. UGC 7321, with an even lower SFR, also has an \HI\ halo as shown by \cite{matthews_wood_2003}.

Which other galactic properties are relevant to the characteristics of gas in halos? What role does cold gas accretion play? The full HALOGAS Survey will provide access to these questions, by providing uniformly sensitive data over a large sample of galaxies, with a broad range in properties such as mass, SFR, and environment.

\acknowledgements{The WSRT is operated by ASTRON (Netherlands Institute for Radio Astronomy) with support from the Netherlands Foundation for Scientific Research (NWO).}

\end{document}